\documentclass[superscriptaddress,aps,prl,twocolumn]{revtex4}
\usepackage{graphicx}
\usepackage{amsmath}
\usepackage{bm}
\usepackage{color}
\begin{document}
\title{An ideal quantum clock and Principle of maximum force}
\author{Yu.\,L.\,Bolotin}
\email{ybolotin@gmail.com}
\author{V.\,A.\,Cherkaskiy}
\email{vcherkaskiy@gmail.com} \affiliation{A.I.Akhiezer Institute
for Theoretical Physics, National Science Center "Kharkov Institute
of Physics and Technology", National Academy of Science of Ukraine,
Akademicheskaya Str. 1, 61108 Kharkov, Ukraine}
\author{A.\,V.\,Tur}
\affiliation{Universit\'{e} de Toulouse [UPS], CNRS, Institut de
Recherche en Astrophysique et Plan\'{e}tologie, 9 avenue du Colonel
Roche, BP 44346, 31028 Toulouse Cedex 4, France}
\author{V.\,V.\,Yanovsky}
\affiliation{Institute for Single Crystals, National Academy of
Science Ukraine, Lenin Ave. 60, 61001 Kharkov, Ukraine}
Ukraine\date{\today}
\begin{abstract}
We show that the space-time uncertainty relation for the quantum clock can be derived from
the maximum force principle.
\end{abstract}
\maketitle Achievement of required accuracy in any quantum
measurement inevitably imposes certain limitations on
characteristics of the device designed to perform it
\cite{Wigner_1957,Hossenfelder_2013}. Let us consider measurement of
short time intervals. All possible methods to measure the time
always involve observation of some physical process. The commonly
known examples are periodic oscillations (all kinds of mechanical
and electronic clocks), stationary flow of liquid (the clepsydra) or
friable (the hourglass) substances, repeated motion of celestial
bodies (the sundial and calenders). The so-called light-clock (a
light signal periodically bouncing between two plane parallel
mirrors, facing each other) plays a principal role in the Special
Relativity. In the present paper we follow
\cite{Burderi_DiSalvo_Iaria_2016} to consider the so-called quantum
clock based on observation of the radioactive decay described by the
equation
\begin{equation}\label{e1}
\frac{dN}{dt}=-\lambda N,
\end{equation}
where $N(t)$ is the current number of radioactive particles in the
sample. Average number of the decayed particles (or nuclei) during
the time interval $\Delta t\ll\lambda^{-1}$ is $\Delta N=\lambda
N\Delta t$. It enables us to measure the time intervals calculating
number of the decayed particles
\begin{equation}\label{e2}
\Delta t=\frac{\Delta N}{\lambda N}.
\end{equation}
Relative error of such time measuring method is
$\varepsilon=(\lambda N\Delta t)^{-1}=1/\sqrt{\Delta N}\le1$. Increasing size of the quantum clock (the number N), one seemingly would gain unlimited improvement in accuracy of the time interval measurement. Burderi et all \cite{Burderi_DiSalvo_Iaria_2016} showed however that it is not true.

Let us represent the measured time interval (\ref{e2}) in the following form
\begin{equation}\label{e3}
\Delta t=\frac1{\varepsilon^2Mc^2}\frac{E_p}{\lambda},
\end{equation}
where $M\equiv m_pN$ is mass of the clock, $m_p$ is mass of the
decaying particle and $E_p=m_pc^2$. It is desirable to get rid of
the factor $E_p/\lambda$ which characterizes the considered quantum
clock.

Being a quantum device, the quantum clock must obey the standard uncertainty relation
\begin{equation}\label{e4}
\Delta E\cdot\Delta t\ge\hbar/2,
\end{equation}
where $\Delta E$ is the maximum accuracy in energy of a quantum system achievable in the measurement process during the time interval $\Delta t$. Evidently, due to the energy conservation law, $\Delta E\le E_p$ and $\Delta t\le1/\lambda$ because $\Delta N\le N$ in (\ref{e2}). Thus $E_p/\lambda\ge\hbar/2$ and from (\ref{e4}) one obtains the lower limit of the quantum clock mass suitable to measure the time interval $\Delta t$ with accuracy $\varepsilon$
\begin{equation}\label{e5}
M\ge\frac\hbar{2\varepsilon^2c^2}\frac1{\Delta t}.
\end{equation}
The result looks encouraging: sufficiently massive clock enables us to measure the short time intervals with the required accuracy. However some questions should be clarified.

In General relativity the time interval is a local characteristic, so in an inhomogeneous gravity field the quantum clock of finite size can measure only averaged (over the clock size) time intervals. One has to minimize the clock size in order to minimize the uncertainty in the time determination. Doing it, one inevitably faces the principal limitation. If the clock radius $R$ (assuming spherical shape of the clock) becomes less than the corresponding gravitational radius $R_g=2MG/c^2$, then we loose possibility to use the clock to measure time, as we cannot any more receive information about the decay products hidden from us behind the event horizon of the corresponding black hole that have formed. The condition $R>R_g$ can be rewritten in the form
\begin{equation}\label{e6}
\frac1M>\frac{2G}{c^2R}.
\end{equation}
Combining (\ref{e6}) with (\ref{e5}), one obtains
\begin{equation}\label{e7}
\Delta t R>\frac1{\varepsilon^2}\frac G{c^4}\hbar.
\end{equation}
Treating $R$ as uncertainty $\Delta r$ in position of the physical object (the clock), which is the basis for the time measurement process, and taking into account that $\varepsilon\le1$, one finally obtains \cite{Burderi_DiSalvo_Iaria_2016}
\begin{equation}\label{e8}
\Delta t \Delta r>\frac G{c^4}\hbar.
\end{equation}
The obtained inequality restricts the possibility to determine temporal and special coordinates of an event with arbitrary accuracy.

Any information about the particular process, which the clock is
based on, is absent in the relation (\ref{e8}), so it suggests that
this relation can be obtained from very general considerations. For
that purpose we use the maximum force principle
\cite{Gibbons_2002,Schiller_2006,Barrow_Gibbons_2014}, which states
that the tension or force between two bodies cannot exceed the value
\begin{equation}\label{e9}
F_{max}=\frac{c^4}{4G}\approx3.25\times10^{43}N.
\end{equation}
This limit does not depend on nature of the forces and is valid for
gravitational, electromagnetic, nuclear or any other forces. It can
be shown that the maximum force principle can be derived from the
holographic principle and vice versa \cite{Bolotin_Cherkaskiy_2015}.

Existence of the maximum force is generally speaking a principle (postulate), but it is quite clear how it comes to being. Let us consider the attractive gravitational force between two masses $M_1$ and $M_2$, separated by the distance $D$ (we use the Newtonnian approximation for simplicity). The forces between them are
\begin{equation}\label{e10}
F= G\frac{M_1M_2}{D^2} = \left(\frac{GM_1}{c^2D}\right) \left(\frac{GM_2}{c^2D}\right)\frac{c^4}G.
\end{equation}
As \[M_1M_2\le\frac14(M_1+M_2)^2,\] then
\begin{equation}\label{e11}
F\le\left[\frac{(M_1+M_2)G}{c^2D}\right]^2\frac{c^4}{4G}.
\end{equation}
Mutual approaching of the bodies is limited by the condition \[R>R_g=\frac{2MG}{c^2},\] which prohibits creation of the black hole with mass $M=M_1+M_2$. Therefore
\begin{equation}\label{e12}
F\le\frac{c^4}{4G}.
\end{equation}

The surfaces which realize the maximum force (the maximum momentum flow) or the maximum power (the maximum energy flow) are horizons. A horizon appears at any attempt to surpass the force limit. And the horizon prohibits the possibility to surpass the limit.

Let us turn back now to the fundamental relation (\ref{e8}). Using the expression (\ref{e9}) for the limiting force, rewrite (\ref{e8}) in the form
\begin{equation}\label{e13}
\Delta t \Delta r>\frac1{F_{max}}\hbar.
\end{equation}
With the Planck constant $\hbar$ fixed, it is only the limiting force $F_{max}$ which determines the restriction on the quantum clock size. If the force limiting value is absent in the theory, i.e. $F_{max}=\infty$, then $R_g\to0$ and the limitation on the quantum clock size disappears.

In order to derive the relation (\ref{e8}) or (\ref{e13}), use the standard uncertainty relation
\begin{equation}\label{e14}
\Delta x_{min} \Delta p_{max}\ge\frac\hbar2.
\end{equation}
Taking into account that \[F_{max}=\frac{\Delta p_{max}}{\Delta x_{min}},\] one immediately obtains that the minimum size of the clock, required to measure time intervals $\Delta t_{min}$, obeys the restriction
\begin{equation}\label{e15}
\Delta x_{min} \Delta t_{min}\ge\frac\hbar{F_{max}},
\end{equation}
which is identical to (\ref{e13}).

To conclude, we would like to stress that the main reason for the restriction on the quantum clock size is the requirement $R>R_g$, equivalent to the condition which prohibits formation of the horizon. Therefore it is quite natural that the righthand side of the relation (\ref{e13}) contains the limiting force value $F_{max}$, which can only be achieved on a horizon.

\end{document}